\documentstyle[aps,pre,epsf,multicol]{revtex}
\begin{document}
\draft
\title{Reconstructed Rough Phases During Surface Growth}
\author{ Chen-Shan Chin and Marcel den Nijs}
\address{ Department of Physics, University of Washington, P.O. Box 351560, \\
Seattle, Washington 98195-1560}
\date{\today}
\maketitle

\begin{abstract}
Flat surface phases are unstable during growth and known to become
rough.  This does not exclude the possibility that surface
reconstruction order persists in rough growing surfaces, in analogy
with so-called equilibrium reconstructed rough phases.  We investigate
this in the context of KPZ type dynamics, using the restricted solid
on solid model with negative mono-atomic step energies.  Long range
reconstruction order is strictly speaking absent in the thermodynamic
limit, but the reconstruction domain walls become trapped at surface
ridge lines, and the reconstruction order parameter fluctuates
critically with the KPZ dynamic exponent at finite but large length
scales.
\end{abstract}

\begin{multicols}{2}
\narrowtext

Equilibrium surface phase transitions have been studied in great
detail during recent decades.  Various types of critical behaviors
emerged both in theoretical models and experimental
systems~\cite{surfrevs}.  Dynamic non-equilibrium aspects are still
less well understood, in particular whether any of those equilibrium
transitions persists in the stationary state properties of growing
interfaces.  For example, equilibrium crystal surfaces undergo
roughening transitions from macroscopic flat to rough structures,
while growing surfaces are believed to be rough under all
circumstances~\cite{Nozieres,Villain}, as confirmed by numerous
studies of, e.g., KPZ type dynamics~\cite{KPZ} and other dynamic
universality classes.  Still, it is custom to identify distinct growth
regimes within this rough phase.  So-called step-flow type
layer-by-layer growth is an example.  At low temperatures, well below
$T_R$ where the equilibrium surface roughens, a new terrace is
nucleated with an exponential small probability.  The time scale at
which this nucleus grows by particle adhesion at its edge into a
macroscopic domain is much shorter than the time scale at which a
nucleus for the next layer appears on top of this terrace.  So at
small enough length scales the surface looks flat and seems to grow
layer-by-layer.  In this letter we introduce a similar transient
phenomenon for surface reconstruction order in growing rough phases.

Surface reconstruction is usually associated with flat interfaces.
However, roughness not necessarily destroys the reconstruction order.
Equilibrium reconstructed rough (RR) phases are known to exist for
so-called misplacement type reconstruction~\cite{MdN-Pt110,MdN-King}.
The compatibility of surface reconstruction with surface roughness
depends on intricate topological aspects.  For example, in missing row
reconstructed FCC (110) facets the reconstruction couples
strongly to the surface roughness such that surface roughening
simultaneously destroys the reconstruction order
~\cite{MdN-Pt110}. For other symmetries, like simple cubic (SC) (110)
missing row reconstruction, they decouple.

The RR order parameter must be formulated with care ~\cite{MdN-King}.
In flat SC missing row structures for example, the reconstruction
order can be formulated in two ways, which seem equivalent at first,
but at the roughening transition only one of them vanishes.  One
formulation keeps track of whether the even or odd rows are on top.
The other one measures it by the (striped) antiferromagnetic ordering
of the parity type Ising variables $S_r= \exp(i \pi h_r)$, with $h_r$ 
the surface height.  Step excitations in this surface
belong to two distinct topological sets.  One couples only to the
first and the other to the second order parameter.  At the roughening
transition only the free energy of the cheapest type of steps goes to
zero.  Simultaneously, the reconstruction order parameter that couples
to it vanishes, but the other type of order persists inside the rough
phase.  Only the parity type order parameter is readily observable in,
e.g., x-ray diffraction. Therefore, the reconstruction order of the
rough phase can go unnoticed. For more details, see
ref.~\cite{MdN-King}.

It is conceivable that reconstruction order persists in growing
surfaces.  Imagine a two dimensional (2D) lattice with on each site an
height variable and a spin degree of freedom (representing the
reconstruction order).  This leads to two coupled master equations,
one for surface growth, e.g., KPZ type dynamics, and another one for
the reconstruction order, e.g., Glauber type Ising dynamics.  The
local growth probability varies with the Ising configuration and the
Ising spin flip probabilities are affected by the local surface height
profile.  Are these couplings relevant or irrelevant?  If irrelevant,
the KPZ sector evolves into the stationary non-equilibrium KPZ type
state and the Ising degrees of freedom reach the Gibbs equilibrium
state (even while the surface is growing).  Coupled master equations
of this type have been studied recently in the context of specific 1D
growth models.  Those display strong coupling between the Ising and
roughness degrees of freedom, such as growth being pinned down by
Ising domain walls~\cite{Drossel,Kotrla,Noh}.  We observe a different
type of strong coupling.
 
The 2D restricted solid on solid (RSOS) is one of the work horses of
surface physics research.  Nearest neighbor heights differ by at most
one, $dh=0,\pm1$. 
\begin{equation}
E =  \frac{1}{2} K \sum_{\left<r, r^{\prime}\right>} (h_r-h_{r^{\prime}})^2
\end{equation}
with only nearest neighbor interactions, and 
dimensionless units, $K=J/k_{\rm B}T$.  The $J>0$ side of the
phase diagram contains a conventional equilibrium surface roughening
transition~\cite{MdN-RSOS} and the growth model version has been
studied extensively for $J>0$ as well~\cite{KPZ,KK,RSOS,CCS-MdN}.

For $J<0$, the model contains one of the simplest examples of a
reconstructed rough phase~\cite{MdN-RSOS}, and is probably the most
compact formulation of the coupling between Ising and surface degrees
of freedom.  The $dh=\pm1$ steps are more favorable than flat $dh=0$
segments.  At zero temperature, the $dh=0$ states are frozen out, and
the model reduces to the so-called body centered solid on solid
(BCSOS) model.  Its surface is rough, but since nearest neighbor
heights must differ by one, all heights on one sublattice must be even
and odd at the other, or the other way around.  This two-fold
degeneracy represents a checker board type RR phase.
The staggered magnetization $m$, defined in terms of the parity spin
type variables $S_i = \exp(i\pi h_r)$, is non-zero.

The $dh=0$ excitations appear at $T>0$. They form closed loops, and
behave like Ising type domain walls. The reconstruction order $m$
changes sign across such loops.  Their sizes diverge at the
equilibrium deconstruction transition $K_{\rm c}=-0.9630$~\cite{MdN-RSOS}.  
It was found that the Ising and roughness variables only couple weakly, 
i.e., that all reconstruction aspects of the transition follow conventional
Ising critical exponents.  Moreover, the thermodynamic singularities
in the Ising sector only affect the temperature dependence of the
surface roughness parameter $K_{\rm G}$ inside the rough phase.  The
latter is defined in terms of the logarithmic divergence of the
height-height correlations,
\begin{equation}
\langle (h_{r+r_0}-h_{r_0})^2 \rangle \simeq (\pi K_{\rm G})^{-1} \ln(r).
\end{equation}
The continuum limit confirms this weak coupling.
The decoupling point of the Gaussian and Ising degrees of freedom
is there a stable renormalization type fixed point~\cite{MdN-King}.

We present only our Monte Carlo (MC) simulation results in the
far from equilibrium limit where evaporation becomes forbidden.
The results look similar closer to equilibrium, but
crossover scaling phenomena make a quantitative analysis more difficult
(as expected).  During the MC simulation we keep a list of active
sites, where particles can deposit without violating the RSOS
condition.  They are grouped in $j=1,\cdots,5$ sets,
according to the only five distinct energy changes $\Delta E_j$ 
that can occur during deposition.  
First we preselect one of those 5 sets, with probability 
$(p_j N_j)/(\sum_j p_j N_j)$, where $p_j =\min(1,e^{-\Delta E_j})$ 
and $N_j$ is the number sites of type $j$.
Next, a particle is randomly deposited at one 
of the sites in that specific set $j$ .
Rejection free procedures like this upset the proper flow
of time.  We need it, because  the Metropolis dynamics
slows down at low temperatures
due to an high rejection rate and lack of active sites.
To restore proper time, we increase the MC time
during each update step by $1/p\times1/N_j$. 
This reproduces the correct value for the KPZ dynamic exponent
$z\simeq 1.6\pm 0.1$ at all temperatures.

\begin{figure}
\centerline{\epsfxsize=8cm \epsfbox{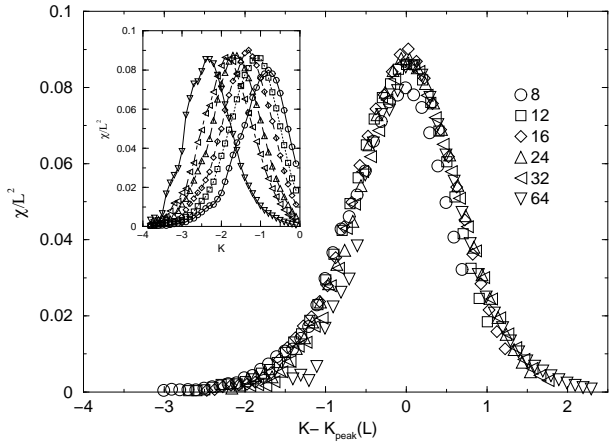}}
\vskip 10 true pt
\caption{
Reconstruction order parameter susceptibility $\chi$ as function 
of temperature at system sizes $L=8$-$64$.
The data collapses onto a single curve 
by the shift $K^\prime = K-K_{\rm{peak}}(L)$, with  
$K_{\rm{peak}}(L)=-0.77\ln(L/2.2)$.}
\label{susClps}
\end{figure}

Fig.\ref{susClps} shows the susceptibility, 
$\chi = L^2(\langle m^2 \rangle - \langle | m | \rangle^2)$\cite{MC-suscep},
as function of temperature for different system sizes $L^2$.  The
sharpt maxima seem to confirm the existence of a RR phase, but several
features are very different from equilibrium.  The height and width of
$\chi/L^2$ do not scale with $L$.  At conventional equilibrium
transitions, the peak height decreases as $\chi/L^2\sim
L^{\gamma/\nu-2}\sim L^{-1/4}$.  Moreover, the peak position does not
converge to a specific critical point. Instead it shifts
logarithmically, as $K_{\rm peak}(L)\simeq-A\ln(L/L_0)$ with $A=
0.77\pm 0.05$ and $L_0= 2.2\pm 0.2$.

\begin{figure}
\centerline
{\epsfxsize=8 cm \epsfbox{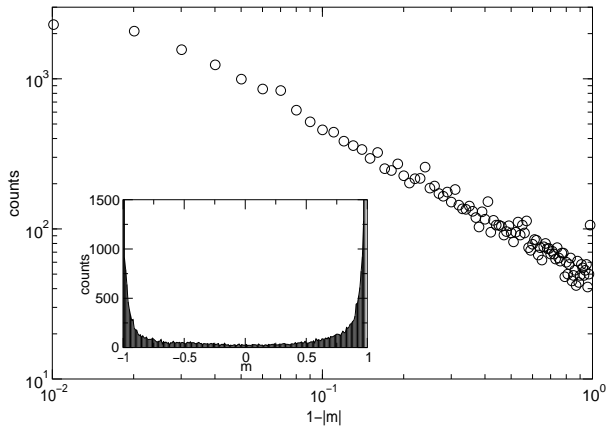}}
\vskip 10 true pt
\caption{
Histogram (shown as insert) of the reconstruction order parameter, $m$, at
$L=32$ and $K=-3.2$ from
a total of $2^{18}$ data points using $\Delta M=0.01$ as bin width.
The tails about the peaks at $m=\pm 1$ scale as power laws
with exponent $-0.9\pm0.1$.}
\label{mhstgm}
\end{figure}

Next, we monitor the reconstruction order parameter $m$ at $K\ll 0$
as function of time.  It behaves similar as in
conventional spontaneously ordered phases, but flip-flops more
frequently than justifiable from finite size effects alone. Moreover,
the fluctuations in $m$ within each phase are too strong.
Fig.\ref{mhstgm} quantifies this in terms of a histogram of the number
of times a specific value of $m$ appears in a typical time series.
The distribution has two distinct peaks, suggesting the presence of
spontaneously broken reconstruction order, but the tails have a power
law shape instead of the mandatory exponential form.
 
\begin{figure}
\centerline{\epsfxsize = 8 cm \epsfbox{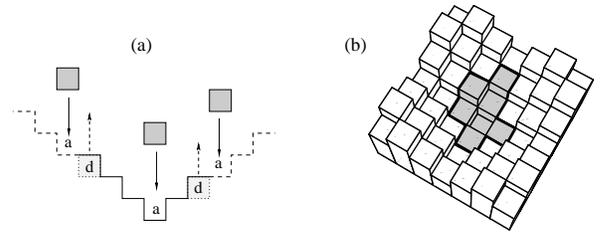}}
\vskip 10 true pt
\caption{ (a) One dimensional cross section of the surface near a valley
with two loop segments.
The sites {\it a} ({\it d\,}) are the only active adsorption (desorption) sites.
The domain walls always move upwards during adsorption. (b) A loop of size
of $l_c$ nucleated at the bottom of the local valley. Gray and white 
sites have different surface reconstruction parity order.}
\label{1D_dyn}
\end{figure}

\begin{figure}
\centerline{\epsfxsize = 8 cm \epsfbox{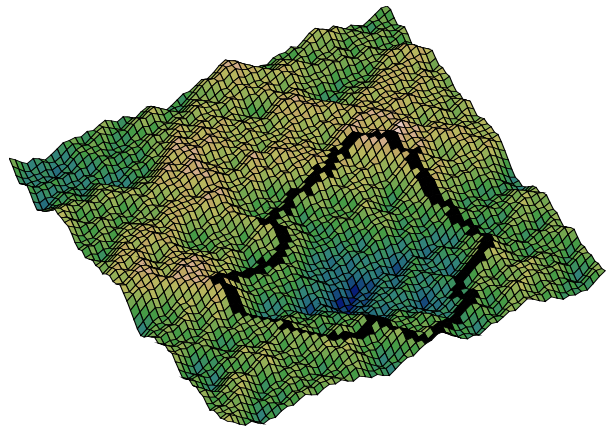}}
\vskip 10 true pt
\caption{
A MC surface configuration (at $K=-3.3$ and $L = 64$)
with a macroscopic loop (the dark line) trapped at a surface ridge line.}
\label{Surface}
\end{figure}

The above observations suggest the existence of quasi-critical
reconstructed rough behavior at low temperatures.  The origins of this
can be traced to the following loop dynamics.  Fig.(\ref{1D_dyn}a) can
be interpreted as a configuration in the 1D version of our model, or a
cross-section of the 2D surface.  It shows a domain of opposite
reconstruction inside an otherwise perfectly reconstructed rough
configuration.  The two flat segments are the domain walls.  In
equilibrium they move apart/towards each other with equal probability
because deposition/evaporation are equally likely.  However, in the
presence of a growth bias, the defects move more likely upwards than
downwards.  The walls grow in size and move up-hill until they get
trapped at the top of the ridges.  Fig.\ref{Surface} shows a trapped
loop in an actual low temperature MC configuration.  Once pinned at
the ridge line, the loops are slaved to the fluctuations of the
roughness degrees of freedom.  Since the surface fluctuations are
scale invariant (KPZ type in our model), the reconstruction order
parameter fluctuates critically with a power law distribution.  Each
loop has to follow this dance until a new loop nucleates out of the
valley and annihilates it, or when the encircled terrain happens to
shrink to zero (fills-up) by surface growth fluctuations.

The nucleation of loops takes place in local valleys.  Consider a deep
valley in a perfect BCSOS type RR surface configuration, like in
Fig.\ref{Surface}, at low temperatures.  The probability for
deposition of one particle at the bottom of the valley is equal to $p=
L^{-2}\exp(2K)$.  The next event in this local area can destroy the
elementary loop (by deposition at the same site with probability
$p=L^{-2}$) or widen it (by deposition next to it).  Annihilation
events return us to a perfect BCSOS surface that has grown by one
vertical $2\times1$ brick. The elementary growth events in the BCSOS
model are direct depositions of such bricks.  In the RSOS model, this
process requires an intermediate elementary loop excitation state.
This implies that the time clock in the RSOS model runs slower by a
factor $r=\exp(2K) ( 1 + 4 \exp(K) + ... )$ for $K\ll 0$.  This is the
origin of the afore mentioned slowing down of the dynamics for $K\ll
0$.  In the following discussion we measure time in BCSOS units.

Loop fluctuations and surface growth events remain entangled up to a
length scale of about $l_c^2\sim 6$ (Fig.(\ref{1D_dyn}b)).  The
annihilation of a loop larger than $l_c$ requires the nucleation of a
distinct new loop from the valley bottom.  The probability for that is
much smaller than for particle depositions at the loop itself, which
widen the loop and make it rise until it becomes trapped on a ridge
line.

The time intervals at which new a macroscopic trapped domain of
opposite reconstruction order emerges out of a valley is independent
on the size of the enclosed area.  Numerically we find $\tau_n\sim
\exp(- \alpha K)$, with $\alpha=3.0\pm 0.1$, independent of loop size.
$\tau_n$ is of the same order of magnitude as simple estimates for the
nucleation time of a loop of size $l_c$ ($\tau\sim \exp(-4K)$ in BCSOS
time).  This part of the process is the limiting factor.  The second
part, in which the loop grows into a macroscopic trapped object, takes
much less time.  Positional entropy does not renormalize $\tau_n$
either.  Rough surfaces are scale invariant which means that the
notion of valley varies with scale.  The loop enclosed landscape
contains many sub-valleys and sub-hills, and maybe even an high
mountain.  However, only the deepest valley bottom acts as nucleation
site (at low temperatures) because loops nucleated in higher sub
valleys become trapped on sub ridges, and such mountain lake loops can
not grow without additional (rare) nucleation events.

\begin{figure}
\centerline{\epsfxsize = 8 cm \epsfbox{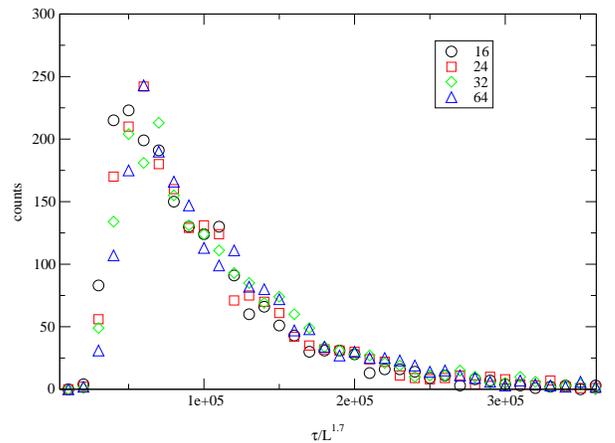}}
\vskip 10 true pt
\caption{ 
Histogram of the decay time of a trapped loop.  
The data collapses onto a single curve by rescaling time 
by a factor $L^{1.7}$.}
\label{timeHis}
\end{figure}

After being trapped on a ridge line, the loop must follow the growth
fluctuations of the surface.  Valleys grow and shrink (without bias),
and fill-up and merge.  $\tau_z$ is the life time of a trapped loop of
size $L$ on a ridge line subject to surface fluctuations only.  We
expect this time to scale as a power law, $\tau_z\sim L^z$, with $z$
the dynamic exponent of the surface roughness degrees of freedom (KPZ
like in our model).  To test this, we measure the decay times of large
macroscopic defect loops (or order $L$) as function of lattice size
$L$.  The data in Fig.(\ref{timeHis}) collapse indeed on one universal
curve after rescaling time by $\tau_z \approx L^z$.  The collapse fits
best at $z=1.7\pm0.1$ (in BCSOS time units), which is consistent with
the KPZ exponent $z=1.6\pm0.1$.

The above analysis presumes that the nucleation time scale $\tau_n\sim
\exp(-\alpha K)$ is larger than the surface growth time scale,
$\tau_z\sim L^z$.  This is valid only well below the equilibrium RR
transition temperature, and only at length scales smaller than, $R_c
\sim \exp(\frac{\alpha}{z} K)$, where loops of size $L$ are being
nucleated infrequently compared to the time scale at which surface
growth washes out surface features of size $L$.  The surface appears
as reconstructed rough for $L<R_c$.  Moreover, the reconstruction
order parameter appears to be fluctuating in a critical manner, since
the loops are trapped to the ridge lines, and are slaved by the
surface fluctuations. So for example, if it were possible to perform
x-ray diffraction from a growing interface, one would observe not only 
power law shaped peaks associated with the surface roughness. At
temperatures where $R_c$ becomes larger than the coherence length of
the surface, additional power law shaped (critical) reconstruction 
diffraction peaks will appear.

At length scales larger than $R_c$, the surface appears as
unreconstructed rough. Loops at that large size die by nucleation of new
loops instead of KPZ surface fluctuations, and they are not 
trapped anymore, because loop segments of can hop across 
sub valleys of size $R>R_c$  near the ridge line by means of nucleation 
of new loops in those mountain valleys.
The peak in the susceptibility, see Fig.\ref{susClps}, reflects this
crossover length $R_c$. Recall that the peak shifts as
$K=-\frac{z}{\alpha}\ln(L/L_0)$.  By setting $\tau_n=\tau_z$ we obtain
the same logarithmic behavior. $A=z/\alpha$ is too small by about
50\%, but this is not a surprise because higher order processes renormalize 
these two time scales near $R_c$.

In conclusion, reconstructed rough phases are absent during growth in
a strict thermodynamic limit sense, but at a more local, and still
large length scales (at low temperatures) the surface grows as if it
is reconstructed with critical fluctuations in the reconstruction
order parameter.  Trapping of the loops to the surface degrees of
freedom at the ridge lines, lies at the core of this.  This behavior
is different from recent results for 1D models with KPZ and Ising type
coupled degrees of freedom.  There, e.g., the Ising defects become
trapped in valleys and canyons and pin-down the
growth~\cite{Drossel,Kotrla}.  We expect to observe crossover to
similar structures in 2D by adding more interactions in our model and
thus vary the local growth rates.~\cite{Chin}.  This research is
supported by the National Science Foundation under grant DMR-9985806.

\end{multicols}

\begin{references}
\bibitem{surfrevs}
{\em The Chemical Physics of
Solid Surfaces and Heterogeneous Catalysis}, Vol. 7 edited by D.~King
(Elsevier, Amsterdam, 1994).
\bibitem{Nozieres}
P.~Nozi\`{e}res and F.~Gallet, J.~Phys.(Paris) {\bf 48}, 353(1987).
\bibitem{Villain}
A.~Pimpinelli and J.~Villain,{\it Physics of Crystal Growth} (Cambridge University Press, 1997).
\bibitem{KPZ} 
M.~Kardar, G.~Parisi, and Y.C.~Zhang, Phys.~Rev.~Lett. {\bf 56}, 889 (1986); 
T.~Halpin-Healy and Y.C.~Zhang, Phys.~Rep.~{\bf 254}, 215 (1995).
\bibitem{MdN-Pt110} M.~den Nijs, Phys.~Rev.~B {\bf 46}, 10386(1992). 
\bibitem{MdN-King} 
M.~den Nijs, chapter 4 in ref.1.
\bibitem{Drossel} B.~Drossel, and M.~Kardar, cond-mat/0002032.
\bibitem{Kotrla} M.~Kotrla, and M.~Predota, Europhys.~Lett. {\bf 39}, 251 (1997); 
M.~Kotrla, F.~Slanina and M.~Predota, Phys.~Rev.~B {\bf 58}, 10003 (1998).
\bibitem{Noh} J.D.~Noh, H.~Park, and M.~den Nijs, Phys.~Rev.~Lett., in press. 
\bibitem{MdN-RSOS} M.~den Nijs, J.~Phys.~A {\bf 18}, L549 (1985).
\bibitem{KK} J.M.~Kim and J.M~. Kosterlitz, Phys.~Rev.~Lett.{\bf 62}, 2289 (1989).  
\bibitem{RSOS} J.G.~Amar and F.~Family, Phys.~Rev.~Lett.{\bf 64}, 543 (1990), 
{\it ibid.} {\bf 64}, 2334 (1990);
J.~Krug and H.~Spohn {\it ibid.} {\bf 64}, 2332 (1990);
J.~Kim, T.~Ala-Nissila and J.M.~Kosterlitz {\it ibid.} {\bf 64}, 2333 (1990).
\bibitem{CCS-MdN} C.S.~Chin  and M.~den Nijs, Phys.~Rev.~E.~{\bf 59} , 2633-2641 (1999).
\bibitem{MC-suscep} K.~Binder, and D.~W.~Heermann, 
{\em Monte Carlo Simulation in Statistical Physics} (Springer-Verlag, Heidelberg, 1997).
\bibitem{Chin} C.S.~Chin, in progress.
\end{references}
\end{document}